\documentclass[twocolumn, titlepage, superscriptaddress, amsmath, amssymb, 10pt]{revtex4-2}

\usepackage{mathtools}        
\usepackage{bbm}
\usepackage{float}            
\usepackage{graphicx}         
\usepackage{color}
\usepackage[dvipsnames]{xcolor}
\usepackage[caption=false]{subfig}
\usepackage{physics}
\usepackage{epstopdf}
\usepackage{comment}
\usepackage{hyperref}
\usepackage{array}
\hypersetup{
	linkcolor=blue,
	citecolor=blue,
	filecolor=blue,
	urlcolor=blue,
	colorlinks=true
}

\usepackage{hhline}
\usepackage{multirow}

\usepackage{soul} 
\renewcommand{\hl}[1]{#1}
\renewcommand{\st}[1]{\iffalse{#1}\fi}

\newcommand{\UIUCPHYS}[0]{Department of Physics, University of Illinois at Urbana-Champaign, Urbana, IL 61801, USA}
\newcommand{\USRA}[0]{USRA Research Institute for Advanced Computer Science, Mountain View, CA 94043, USA}
\newcommand{\NASA}[0]{Intelligent Systems Division, NASA Ames Research Center, Moffett Field, CA 94035, USA}
\newcommand{\KBR}[0]{KBR, Inc., NASA Ames Research Center, Moffett Field, CA 94035, USA}

\newcommand{\AMC}[0]{Applied Mathematics and Computational Research Division, Lawrence Berkeley National Lab, Berkeley, CA 94720, USA}
\newcommand{\NERSC}[0]{NERSC, Lawrence Berkeley National Lab, Berkeley, CA 94720, USA}

\begin{document}

\title{A circuit-generated quantum subspace algorithm for the variational quantum eigensolver}

\author{Mark R. Hirsbrunner}\email{hrsbrnn2@illinois.edu}\affiliation{\UIUCPHYS}\affiliation{\USRA}
\author{J. Wayne Mullinax}\affiliation{\KBR}\affiliation{\NASA}
\author{Yizhi Shen}\affiliation{\AMC}
\author{David B. Williams-Young}\affiliation{\AMC}
\author{Katherine Klymko}\affiliation{\NERSC}
\author{Roel Van Beeumen}\affiliation{\AMC}
\author{Norm M. Tubman}\email{norman.m.tubman@nasa.gov}\affiliation{\NASA}

\begin{abstract}
Recent research has shown that wavefunction evolution in real- and imaginary-time can generate quantum subspaces with significant utility for obtaining accurate ground state energies. Inspired by these methods, we propose combining quantum subspace techniques with the variational quantum eigensolver (VQE). In our approach, the parameterized quantum circuit is divided into a series of smaller subcircuits. The sequential application of these subcircuits to an initial state generates a set of wavefunctions that we use as a quantum subspace to obtain high-accuracy groundstate energies. We call this technique the circuit subspace variational quantum eigensolver (CSVQE) algorithm. By benchmarking CSVQE on a range of quantum chemistry problems, we show that it can achieve significant error reduction \hl{in the best case} compared to conventional VQE, particularly for poorly optimized circuits, greatly improving convergence rates. Furthermore, we demonstrate that when applied to circuits trapped at a local minima, CSVQE can produce energies close to the global minimum of the energy landscape, making it a potentially powerful tool for diagnosing local minima.
\end{abstract}

\maketitle

\section{Introduction}
A main goal of quantum algorithms for Hamiltonian simulation is generating high fidelity groundstates for problems that are difficult to simulate classically. Two important classes of promising algorithms for achieving quantum advantage in Hamiltonian simulation are variational quantum algorithms~\cite{Peruzzo:2014:4213,McClean:2016:023023,Stavenger2022,Bassman2022,Magann2023,Luo2022,Huembeli2021,Aram2021,Arrasmith2021,Anastasiou2022,Carrasquilla2021,Martyn2022,Burton2022,Claudino2020,Tang2021,Chamaki2022,Romero2018,Ayral2022,Tilly2022,Cerezo2021,Luo2022,Guerreschi2020,McClean2017,Smelyanskiy2016,Cao2019,Smith2022,Lu2022,Zhao2020,Pathak2022,Kandala2017,Chamaki2022-1,Lin2020,bassman2022free,sqms2022,Gustafson_2023,Unmuth2022,Hug2020,Sherbert2021,Sherbert2022,Traps2022,Oganov2020,Miro2020,Kirby2021,Kirby2019,Li2021,Jahin2022,abid23,sawaya2023,gustafson2024surrogate,rosa2023,Dborin_2022,roy2023efficient,wierichs2020,goh2023liealgebraic,CerveroMartin2023barrenplateausin} and quantum subspace methods~\cite{Hogg:2000:181,Farhi:2014,Harrigan:2021:332,AspuruGuzik:2005:1705,Hauke:2020:054401,Kitaev:1997:1191,Abrams:1997:2586,Abrams:1999:5162,Ang2022,Kremenetski2021-1,Kremenetski2021,Tubman2020,klymkoRealTimeEvolution2021, motta2023subspace, parrish2019quantum, stair2020multireference, shen2023estimating,krem2023,baker2021,francis2022subspace,mejuto2023quantum, bharti_iterative_2021, PhysRevResearch.5.023200, zheng2024unleashedconstrainedoptimizationquantum}.  Variational quantum algorithms are widely studied, especially the variational quantum eigensolver (VQE), but quantum subspace methods are only recently beginning to grow in popularity. Quantum subspace methods have the potential to provide significant convergence improvements, but challenges remain in the efficient generation of these subspaces on quantum hardware.

In this work, we consider the use of VQE circuits as a mean for generating subspaces with accelerated convergence for groundstate energies of Hamiltonians. We draw inspiration for our approach from recent work studying subspaces generated using real- and imaginary-time evolution~\cite{Motta_2019, klymkoRealTimeEvolution2021,  shen2022realtime, epperly2022theory}. We \hl{forgo the use of a Hamiltonian to guide the time evolution, and instead treat the VQE circuit itself as an evolution operator}, in which the application of each gate in the circuit acts \hl{like a discrete evolution of the state, although not a proper real- or imaginary-time evolution.} The initial state fed into the VQE circuit traces out a path through the Hilbert space as it is ``evolved'' by the VQE circuit, and it is from this path that we sample states to generate a subspace. \st{For an optimized VQE circuit, one might informally consider the series of states generated by the circuit evolution to have similar characteristics to imaginary time evolution, as both rapidly evolve an initial state to the groundstate.} By projecting the Hamiltonian for which the VQE circuit is optimized into this circuit-generated subspace, we form a generalized eigenvalue problem for the Hamiltonian. We call this method the circuit subspace variational quantum eigensolver (CSVQE). In principle, the subspace contains more information than the final state output by the VQE circuit, and the goal of this work is to quantify how much advantage can be extracted from this additional information.

\section{Circuit subspace VQE}\label{sec:algorithm}
The inputs to the CSVQE algorithm are a VQE circuit and the Hamiltonian for which the VQE circuit is optimized. The VQE circuit consists of an ordered series of $N$ parameterized quantum gates acting on an initial wavefunction,
\begin{equation}
    \ket{\Psi_{N}} = U_N U_{N-1} \cdots U_2 U_1 \ket{\Psi_0}.
\end{equation}
\hl{Here we define the gates $U_i$ as generic parameterized unitary operators, and in our numerics below they take the form of exponentiated antisymmetric combinations of fermionic excitation and deexcitation operators. For many ansatz it may be more natural to define $U_i$ as elementary quantum gates, and this requires no modification of the algorithm. In this work we take the parameters of the gates to be constants determined by some prior classical optimization procedure. The first step of the CSVQE algorithm is generating} the set of states explored by the circuit as it evolves the initial wavefunction via the sequential application of the quantum gates. Each of these states is obtained by the ordered application of the first $M \leq N$ gates in the circuit to the initial wavefunction,
\begin{equation}
    \ket{\Psi_M} = U_M U_{M-1} \cdots U_2 U_1 \ket{\Psi_0}.
\end{equation}
The subspace of wavefunctions $\left\{\ket{\Psi_i}\right\}$, $i\in [0,N]$, contains the initial wavefunction $\ket{\Psi_0}$, the final state produced by the circuit, and many additional mid-circuit states. 

Next we construct the generalized eigenvalue problem by calculating the Hamiltonian matrix elements and the overlap matrix in this subspace,
\begin{align}
    H_{ij} &= \mel{\Psi_i}{\hat{H}}{\Psi_j} \\
    S_{ij} &= \braket{\Psi_i}{\Psi_j}.
\end{align}
The resulting generalized eigenvalue problem,
\begin{equation}
    \sum_j H_{ij}x_j = E \sum_j S_{ij}x_j,
\end{equation}
can be solved via classical algorithms for approximate eigenvalues of the original Hamiltonian. \hl{We treat the circuit parameters as fixed in the application of the CSVQE algorithm, thus the CSVQE algorithm can be considered a post-processing step to be performed after a standard VQE calculation. The construction and solution of the above generalized eigenvalue problem can, in principle, be integrated into the optimization process of the VQE circuit, but we leave the question of how best to do that to future work.}

We remark that the overlap matrix, $S$, can be rank deficient, as is commonly
encountered in other work on quantum subspace methods~\cite{klymkoRealTimeEvolution2021,fitz2022}. This occurs most often when the mid-circuit states from nearby layers of the circuit have large overlaps with each other. In this work, we address this problem by projecting out singular values of $S$ that are less than some threshold, which we take to be $1\times10^{-10}$.  However, this threshold has to be adjusted in general when accumulating data from noisy hardware, as discussed in previous work~\cite{klymkoRealTimeEvolution2021}. The final wavefunction produced by VQE is limited by the ansatz and initial state chosen and, in most cases, is not capable of producing the exact groundstate. Solving the generalized eigenvalue problem utilizes these subspace states to generate a new groundstate wavefunction that has equal or larger overlap with the true groundstate, and also a lower energy than any individual wavefunction in the subspace.

\begin{figure}
    \centering
    \includegraphics[width=0.95\linewidth]{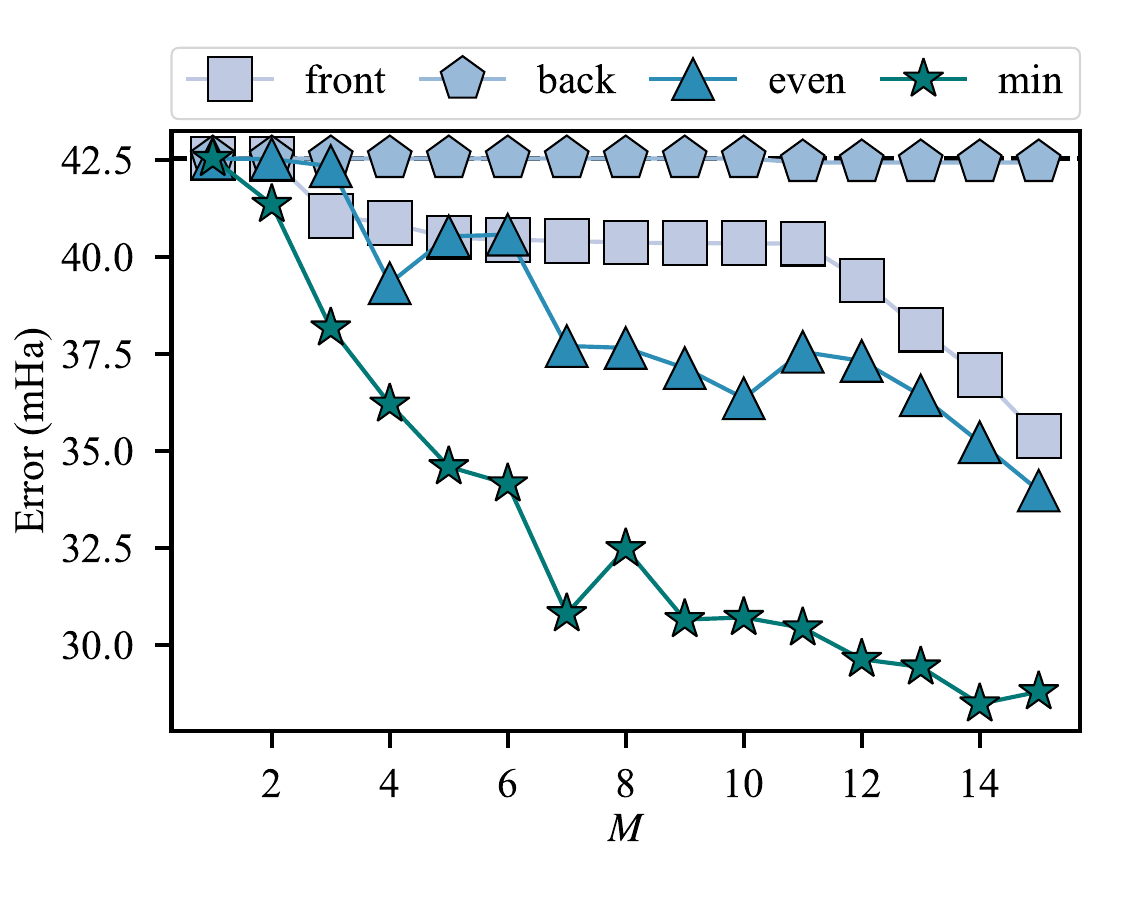}
    \caption[]{The error produced by the CSVQE algorithm when applied to the circuit generated by the first optimization step for $\mathrm{C}_2$ as a function of the number of intermediate states used. The dashed black line is the error in the VQE energy, to which we compare the errors of the CSVQE energies. The gray squares, blue pentagons, and teal triangles mark the error produced by the ``front-loaded'', ``back-loaded'', and ``even'' state selection methods. The green stars mark the lowest error obtained from 20,000 random combinations of mid-circuit states, and $N=149$ for this circuit.}
    \label{fig:C2_energies_vs_n_states}
\end{figure}

In practice, the quantum circuit may contain an enormous number of gates, making it impractical to calculate the matrix elements for the entire circuit-generated subspace. Instead, some subset of $M<N+1$ states from the full subspace must be chosen. However, choosing an optimal subset of states is a challenging problem. To illustrate this for a circuit that has not been fully optimized, in Fig.~\ref{fig:C2_energies_vs_n_states} we plot the groundstate energies obtained by CSVQE for the molecule $\mathrm{C}_2$ using three heuristic approaches and a random selection approach for choosing subsets of the full subspace. We describe these state selection methods below, and describe the details of the circuit used in the following section.

\begin{figure*}[ht!]
    \centering
    \subfloat[$\mathrm{LiH}$]{\label{fig:CSVQEa}\includegraphics[width=0.32\linewidth]{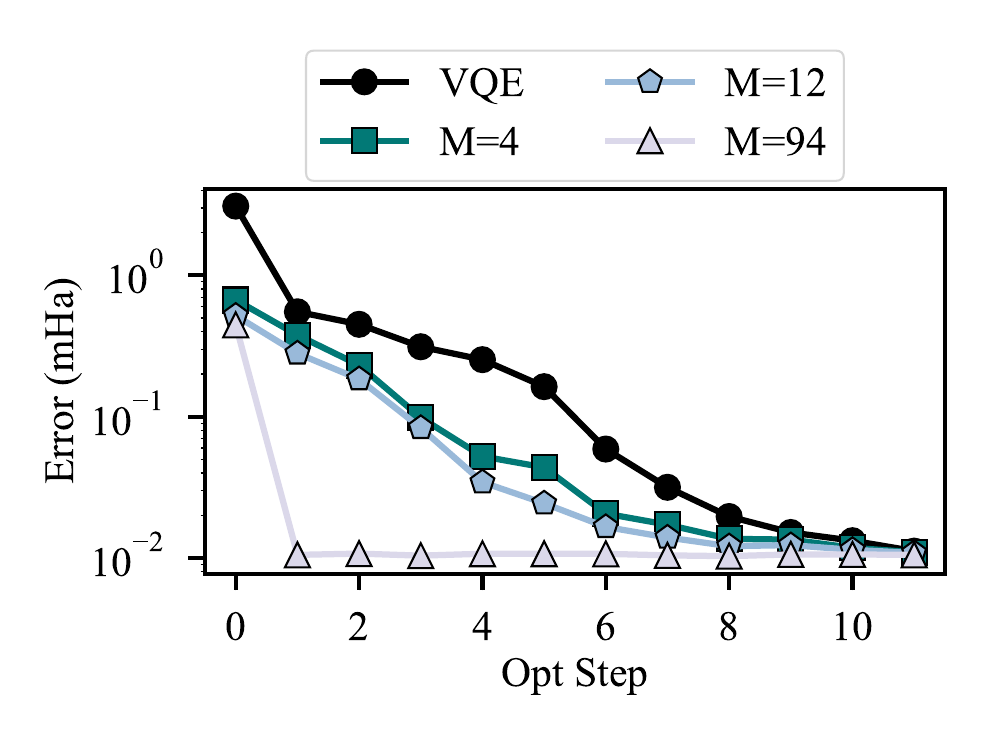}}
    \hfill
    \subfloat[$\mathrm{CH}_4$]{\label{fig:CSVQEb}\includegraphics[width=0.32\linewidth]{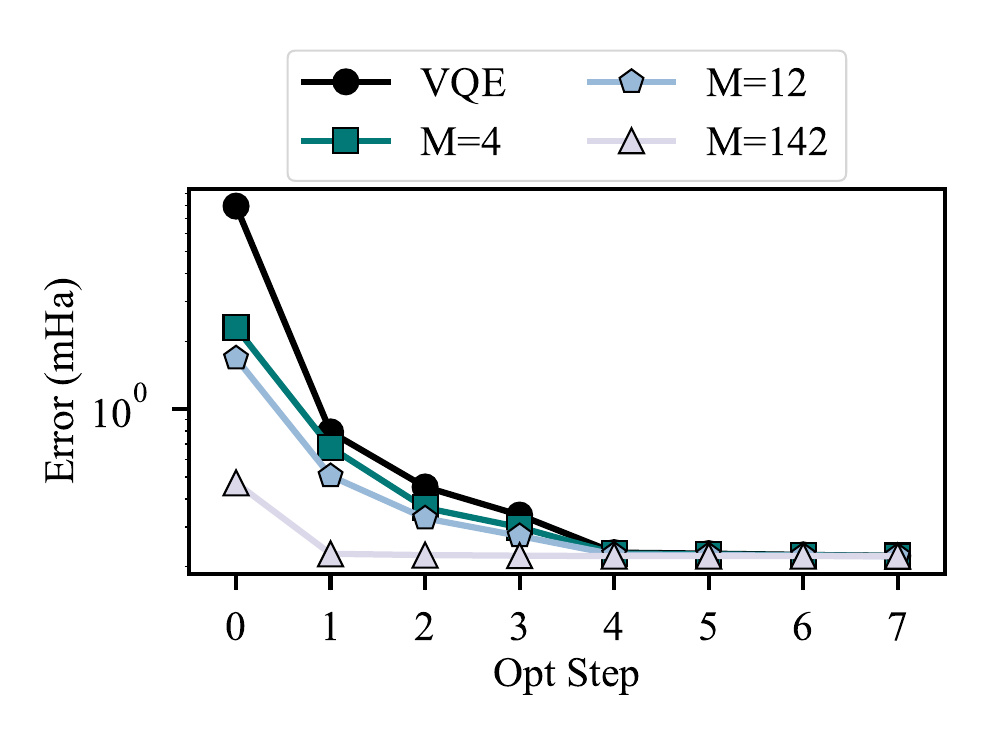}}
    \hfill
    \subfloat[$\mathrm{H}_2\mathrm{O}$]{\label{fig:CSVQEc}\includegraphics[width=0.32\linewidth]{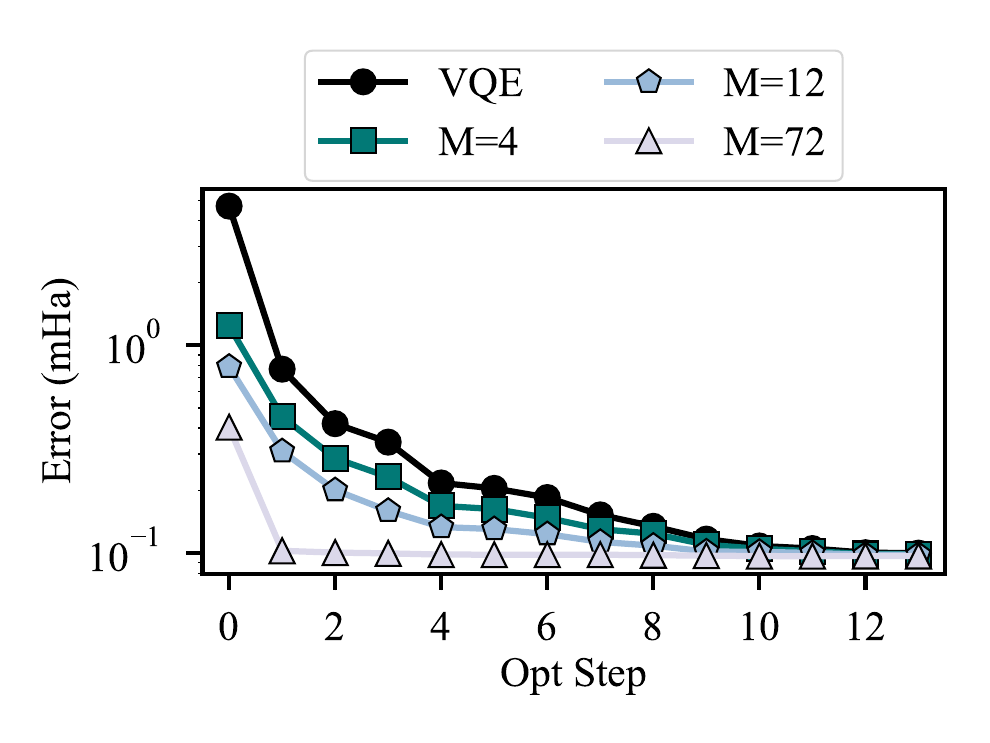}}
    \\
    \subfloat[$\mathrm{NH}_3$]{\label{fig:CSVQEd}\includegraphics[width=0.32\linewidth]{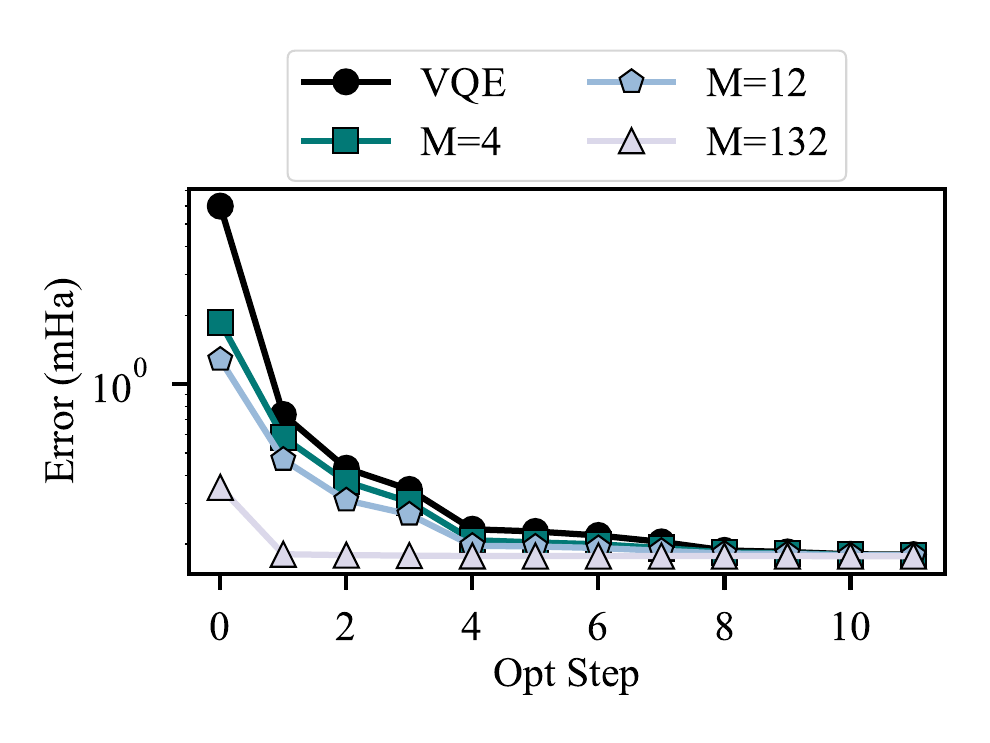}}
    \hfill
    \subfloat[$\mathrm{N}_2$]{\label{fig:CSVQEe}\includegraphics[width=0.32\linewidth]{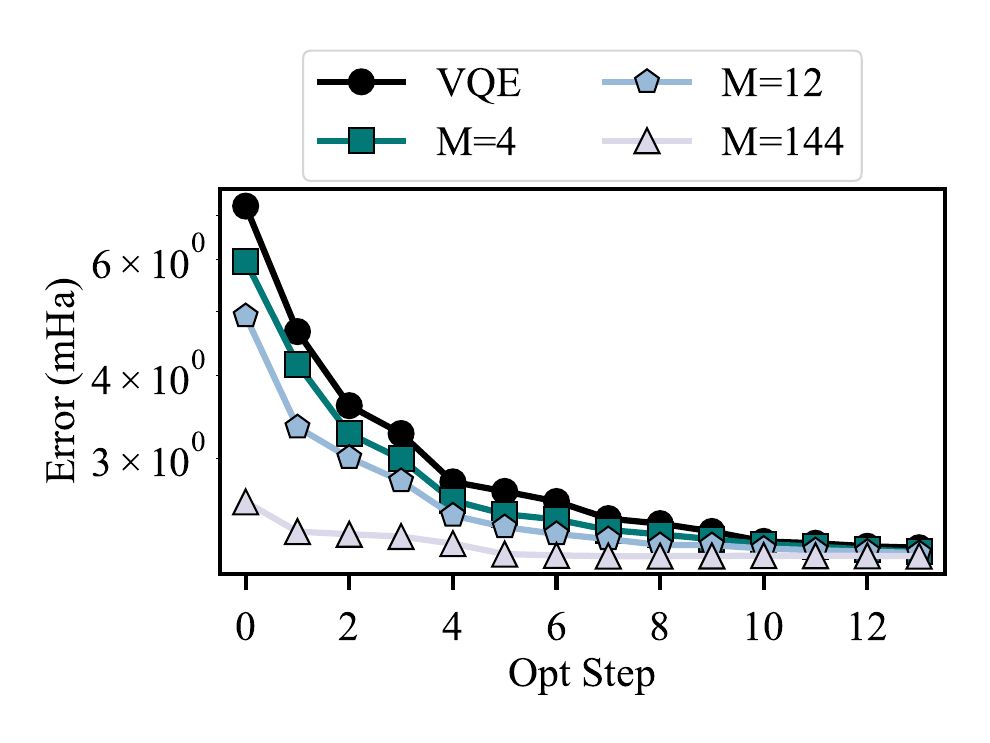}}
    \hfill
    \subfloat[$\mathrm{C}_2$]{\label{fig:CSVQEf}\includegraphics[width=0.32\linewidth]{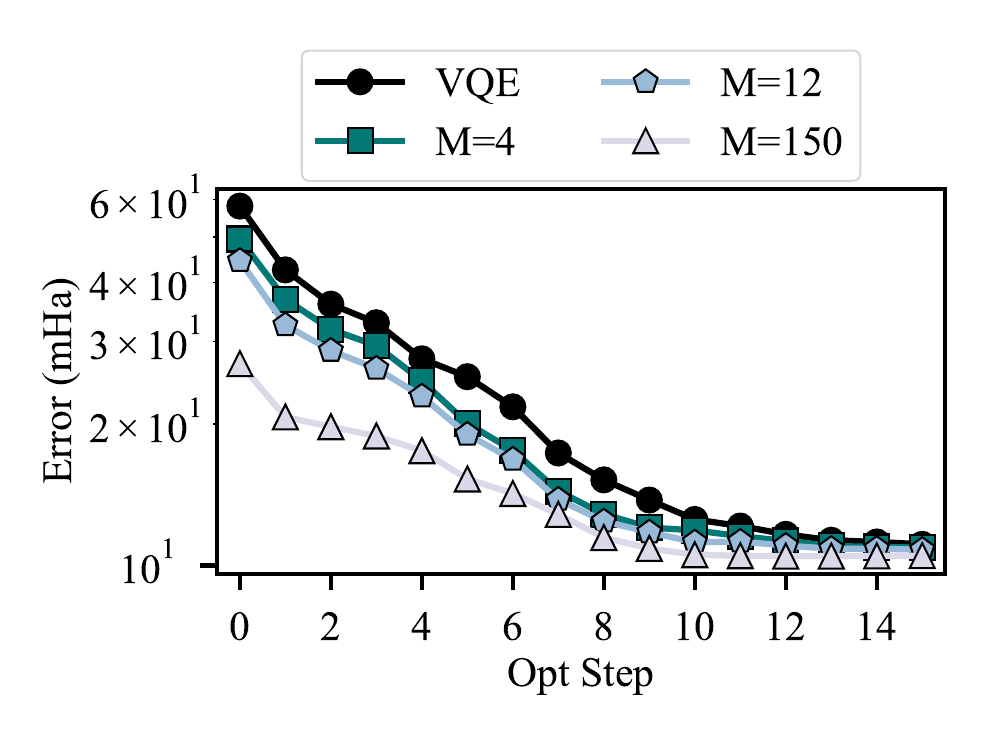}}
    \caption[]{The error in the ground state energy produced by VQE and CSVQE at each step in the classical optimization process for (a) LiH, (b) $\mathrm{CH}_4$, (c) $\mathrm{H}_2\mathrm{O}$, (d) $\mathrm{NH}_3$, (e) $\mathrm{N}_2$, and (f) $\mathrm{C}_2$ using the UCC ansatz. The energies produced by the CSVQE algorithm using four, twelve, and all $(M=N+1)$ mid-circuit states are plotted in green squares, blue pentagons, and gray triangles, respectively. The reported errors represent the optimal values obtained by applying the CSVQE algorithm with 1000 different random combinations of mid-circuit states.}
    \label{fig:CSVQE}
\end{figure*}

The first method we consider is to choose states that are approximately evenly distributed throughout the circuit, e.g. $\left\{\ket{\Psi_i}\right\}$ with $i$ in the set
\begin{equation}
     \left\{0, \lfloor \frac{N}{M - 1} \rfloor, \lfloor \frac{2N}{M - 1} \rfloor, \dots, \lfloor \frac{(M-2)N}{M - 1} \rfloor, N \right\},    
\end{equation}
which we refer to as the ``even'' method. We also consider the ``back-loaded'' method that chooses states only from the end of the circuit,
\begin{equation}
    i \in \left\{N - M + 1, N - M + 2, \dots, N - 1, N \right\},
\end{equation}
and the ``front-loaded'' method, in which states are taken from the beginning of the circuit (also including the final state produced by the full circuit),
\begin{equation}
    i \in \left\{0, 1, \dots, M - 3, M - 2, N \right\}.
\end{equation}
Finally, we consider a random search approach of sampling large numbers of distinct random combinations of states from the subspace. In this way we can try to access nearly optimal energies obtainable with CSVQE, although we note this approach does not scale well and is impractical for use on real quantum hardware. However, in each of these methods, we always include the final state produced by the VQE circuit in the restricted subspace. In doing this we ensure that the CSVQE energy is never higher than the VQE energy, at least within statistical sampling error (shot noise) and biases from noisy hardware.  

As shown by the results in Fig.~\ref{fig:C2_energies_vs_n_states}, the ``even'' and ``front-loaded'' strategies are comparably effective. Both reduce the error but do not quite achieve the accuracy of the best result obtained from 20,000 distinct randomly sampled state combinations. The ``back-loaded'' strategy obtains only minimal advantage over the standard VQE result. \hl{Repeating these calculations for all other systems we study produces similar qualitative behavior for each state-selection strategy.} The intuition behind these results is as follows.  We have ordered the gates in the circuit based on heuristic classical approximations, such that the gates that make the largest changes to the wavefunction are applied first in our construction of the circuit, which we discuss in detail in the following section. States obtained early in the circuit are therefore both the most distinct from the final state produced by the circuit, and also more distinct from their neighboring states than those later in the circuit. As such, they likely provide more utility when solving for the groundstate than the states at the end of the circuit, which have very large overlaps and contain nearly identical information. \st{This is especially true in the case that circuit optimization obtains an excited state with only a small projection onto the groundstate, for which states near the beginning of the circuit likely contain much more useful overlaps with the groundstate.}

\section{Quantum chemistry benchmarking of CSVQE}\label{sec:quantum_chemistry}
In this section we study the effectiveness of the CSVQE algorithm in the context of quantum chemistry, considering a range of molecules and employing the unitary coupled cluster (UCC) ansatz. The UCC ansatz is the exponentiation of the coupled cluster operator $\hat{T}$ acting on the Hartree--Fock reference wave function $\ket{\Psi_0}$,
\begin{gather}
    \ket{\Psi_{\mathrm{UCC}}} = \exp( \hat{T} - \hat{T}^{\dagger})\ket{\Psi_{0}} , \\
    \hat{T} = \sum_{i}^{\mathrm{occ}} \sum_{a}^{\mathrm{vir}} \theta_{i}^{a}\hat{a}_{a}^{\dagger}\hat{a}_{i}
            + \sum_{ij}^{\mathrm{occ}} \sum_{ab}^{\mathrm{vir}} \theta_{ij}^{ab}\hat{a}_{a}^{\dagger} \hat{a}_{b}^{\dagger} \hat{a}_{j} \hat{a}_{i} + \cdots \ .
\end{gather}
The second-quantized creation and annihilation operators $\hat{a}^{\dagger}$ and $\hat{a}$ act on the occupied (indexed by $i,j,\ldots$) and virtual (indexed by $a,b,\ldots$) molecular orbitals in the reference wave function, respectively.

We employ the factorized form of the UCC ansatz,
\begin{equation}
    \label{eq:ansatz}
    \ket{\Psi_{\mathrm{UCC}}} = \prod_{ij\cdots}^{\mathrm{occ}} \prod_{ab\cdots}^{\mathrm{vir}} \hat{U}^{ab\cdots}_{ij\cdots}\ket{\Psi_{0}},
\end{equation}
for which the individual gates are defined as
\begin{equation}
    \label{eq:factor}
    \begin{gathered}
    \hat{U}^{ab\cdots}_{ij\cdots} = \exp(\theta_{ij\cdots}^{ab\cdots}(\hat{a}_{ij\cdots}^{ab\cdots} - \hat{a}_{ab\cdots}^{ij\cdots})) \\
    \hat{a}_{ij\cdots}^{ab\cdots} = \hat{a}^\dagger_a\hat{a}^\dagger_b \dots \hat{a}^{\phantom{\dagger}}_j\hat{a}^{\phantom{\dagger}}_i.
    \end{gathered}
\end{equation}
For simplicity, we include only single- and double-excitation operators, $\hat{U}_i^a$ and $\hat{U}_{ij}^{ab}$. We utilize a computationally efficient implementation of the UCC ansatz that expresses the UCC factors $\hat{U}^{ab\cdots}_{ij\cdots}$ in terms of sines and cosines of the parameters $\theta$~\cite{chenQuantumInspiredAlgorithmFactorized2021}. We choose to order the individual UCC factors based on the magnitude of the initial parameter values ($\abs{\theta}$), which we generate from MP2 calculations using PySCF, such that the factor with the largest coefficient acts first~\cite{sun2018pyscf}. To reduce the computational resources required, we employ a sparse wavefunction circuit simulator designed for simulating the VQE algorithm~\cite{Mullinax2023,hirsbrunner2023}, which is based on techniques developed for classical simulations~\cite{tubman2016,tubman2018-1,levine2020,dbwy2023}. After each UCC factor is applied, we check the number of determinants $N$ in the wave function. If $N$ is greater than the desired number of determinants, $N_{\textrm{WF}}$, we sort the amplitudes by magnitude and discard the determinants with the smallest amplitudes such that only $N_{\textrm{WF}}$ determinants remain in the wave function. We set $N_{\mathrm{WF}}$ to 50,000, which is sufficient to converge the groundstate energy for each molecule tested here.

\begin{figure}[b!]
    \centering
    \subfloat[]{\label{fig:CSVQE_stats_a}\includegraphics[width=0.48\linewidth]{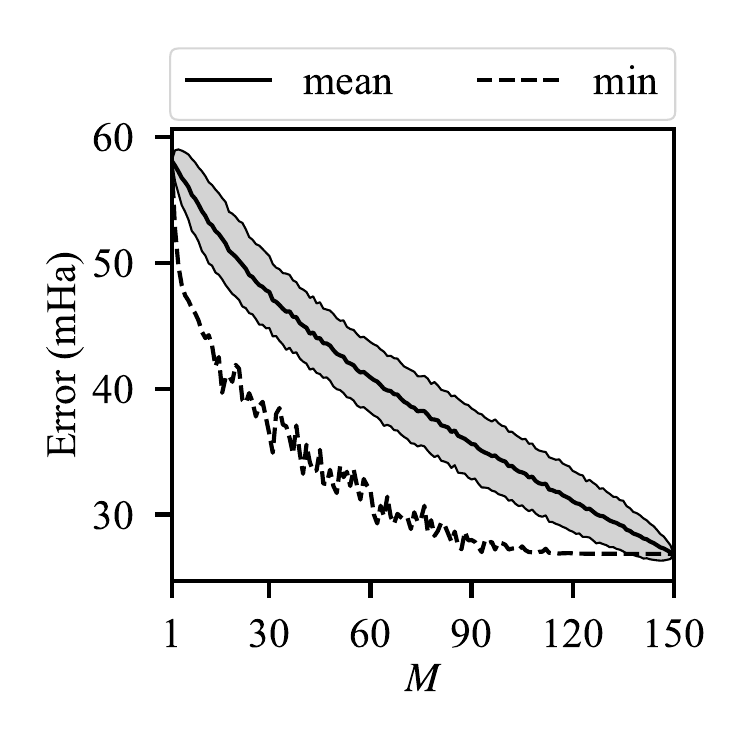}}
    \hfill
    \subfloat[]{\label{fig:CSVQE_stats_b}\includegraphics[width=0.48\linewidth]{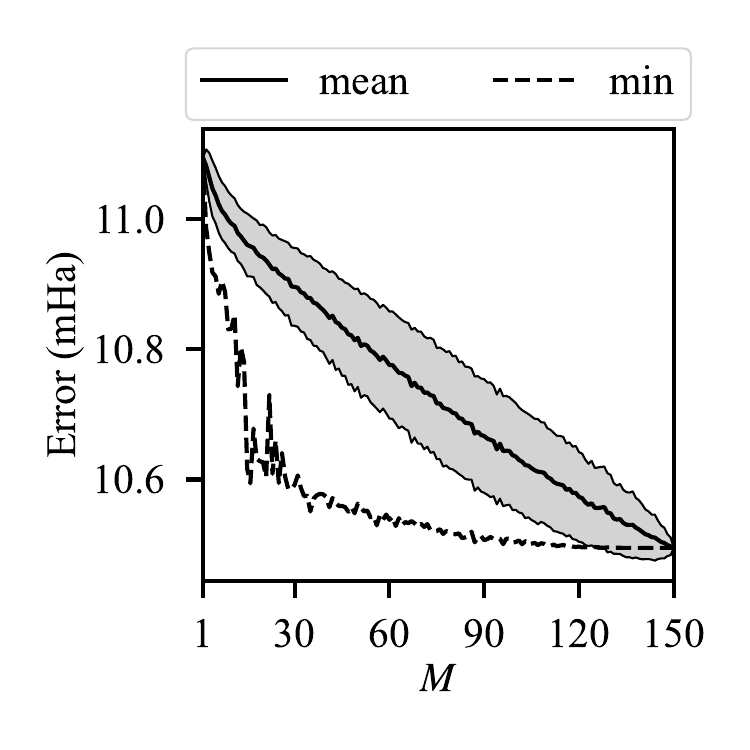}}
    \\
    \subfloat[]{\label{fig:CSVQE_stats_c}\includegraphics[width=0.48\linewidth]{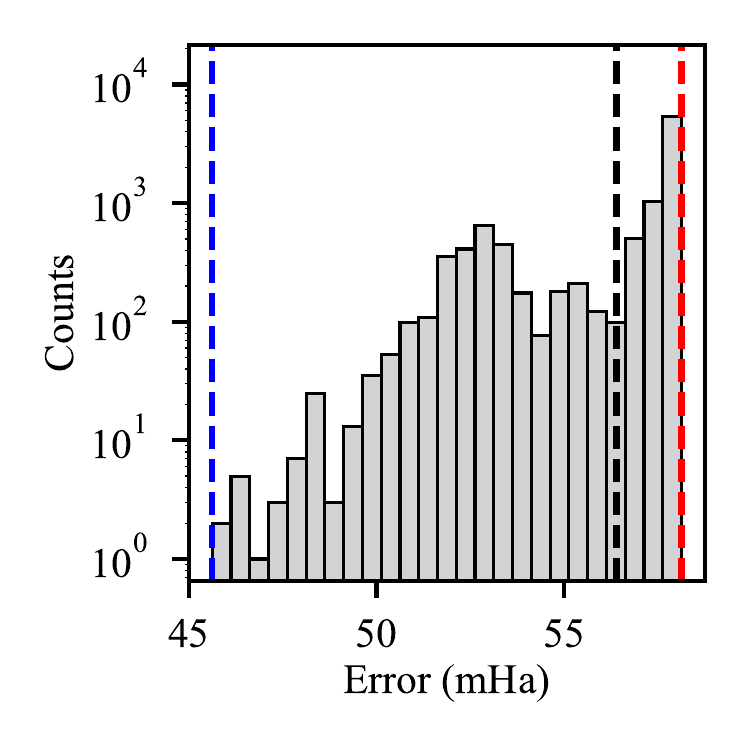}}
    \hfill
    \subfloat[]{\label{fig:CSVQE_stats_d}\includegraphics[width=0.48\linewidth]{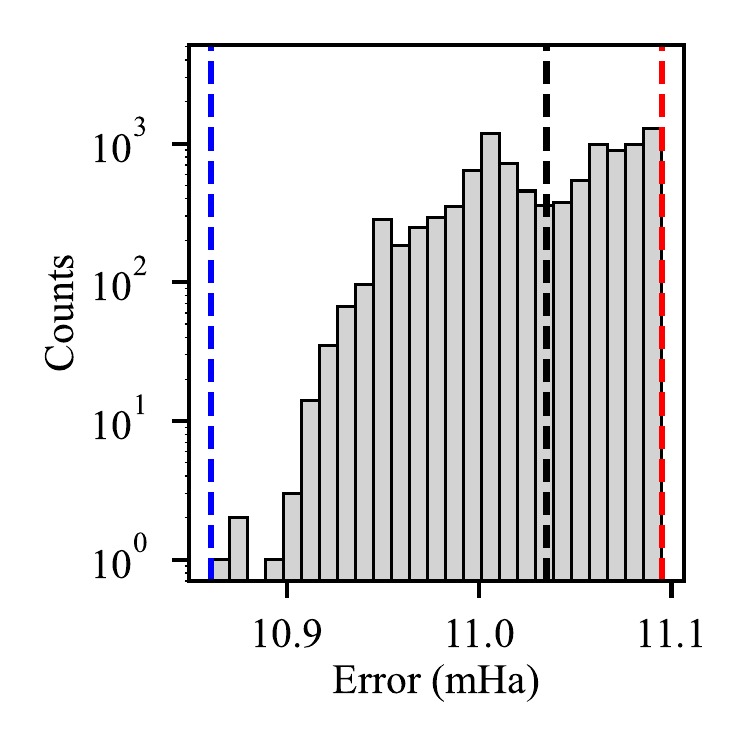}}
    \caption[]{Top: The mean and minimum energies produced by applying CSVQE to the (a) unoptimized and (b) fully optimized $\mathrm{C}_2$ circuits using 1000 random combinations of mid-circuit states, with gray shading indicating the standard deviation. Bottom: Histograms of the energies obtained from 10,000 random combinations of 5 states for the (c) unoptimized and (d) fully optimized $\mathrm{C}_2$ circuits. The red, black, and blue dashed vertical lines denote the standard VQE error, the mean of the CSVQE error distribution, and the minimum CSVQE error, respectively.}
    \label{fig:CSVQE_statistics}
\end{figure}

The results of our benchmarking are summarized in Fig.~\ref{fig:CSVQE}, in which we plot the error of the groundstate energy produced by VQE circuits with and without the CSVQE algorithm at each step in the optimization process for a range of molecule. The details of our benchmark procedure are as follows:

\begin{enumerate}

\item The set of molecules we consider is LiH, $\mathrm{CH}_4$, $\mathrm{H}_2\mathrm{O}$, $\mathrm{NH}_3$, $\mathrm{N}_2$, and $\mathrm{C}_2$. 

\item We use the STO-3G basis set for each molecule and employ experimental geometries from the CCCBDB database~\cite{CCCBDB}.

\item The circuits for each molecule are restricted to include at most fifty doubles operators, choosing the ones with the largest initial parameter values.

\item The circuit parameters are optimized using the Broyden–Fletcher–Goldfarb–Shanno implementation from SciPy, and we store the circuit obtained at each step of the optimization process~\cite{scipy}. 

\item We apply the CSVQE algorithm at each step in the optimization process to study the effectiveness of the algorithm on circuits of varying quality. Rather than deconstructing them into elementary gates, we treat each single and double operator as a single circuit element and extract states after each operator. 

\item We calculate the error relative to the FCI energy and apply the CSVQE algorithm using either four, twelve, or all mid-circuit states. Each data point represents the lowest energy obtained by the CSVQE algorithm from 1000 distinct random combinations of $M$ states chosen from the subspace.

\end{enumerate}

\begin{figure*}[t!]
    \centering
    \subfloat[]{\includegraphics[width=0.32\linewidth]{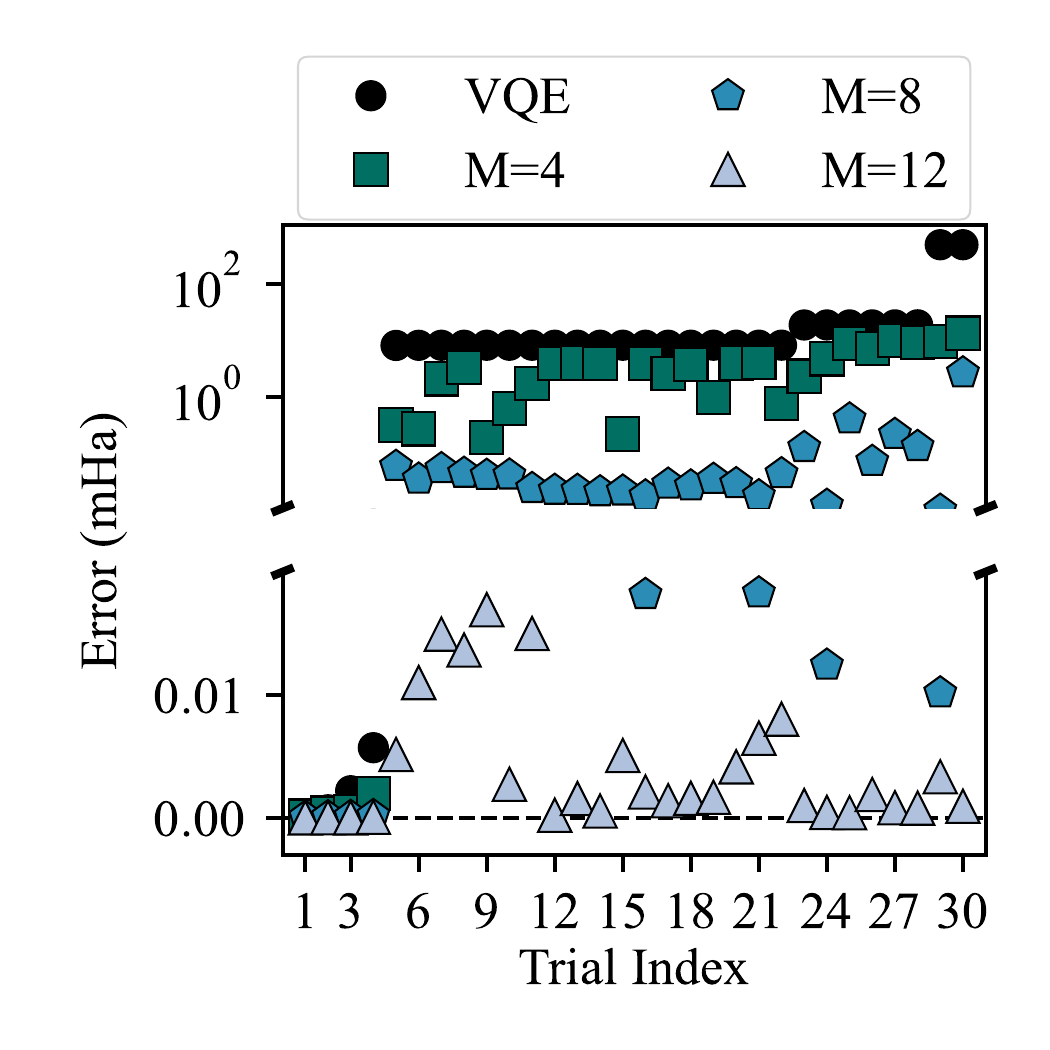}}
    \hfill
    \subfloat[]{\includegraphics[width=0.32\linewidth]{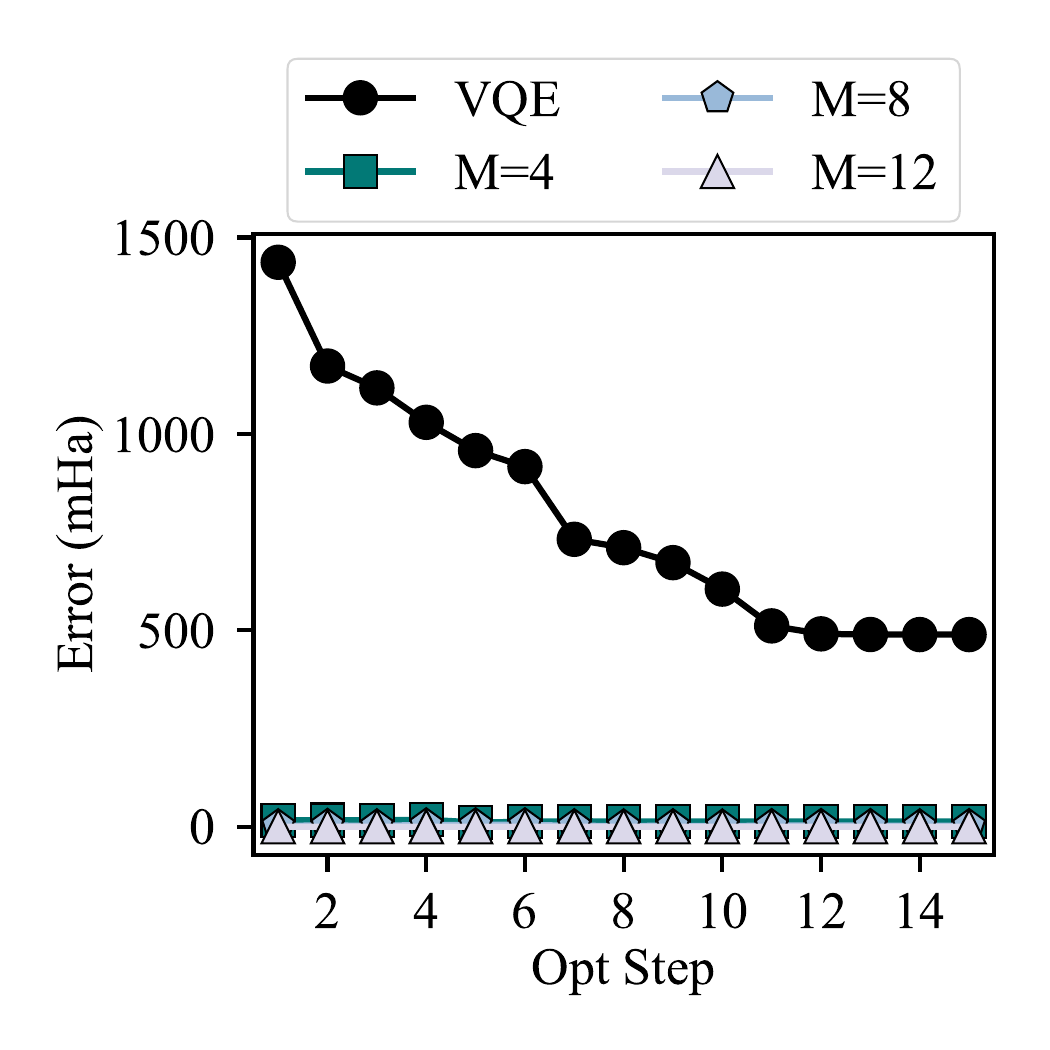}}
    \hfill
    \subfloat[]{\includegraphics[width=0.32\linewidth]{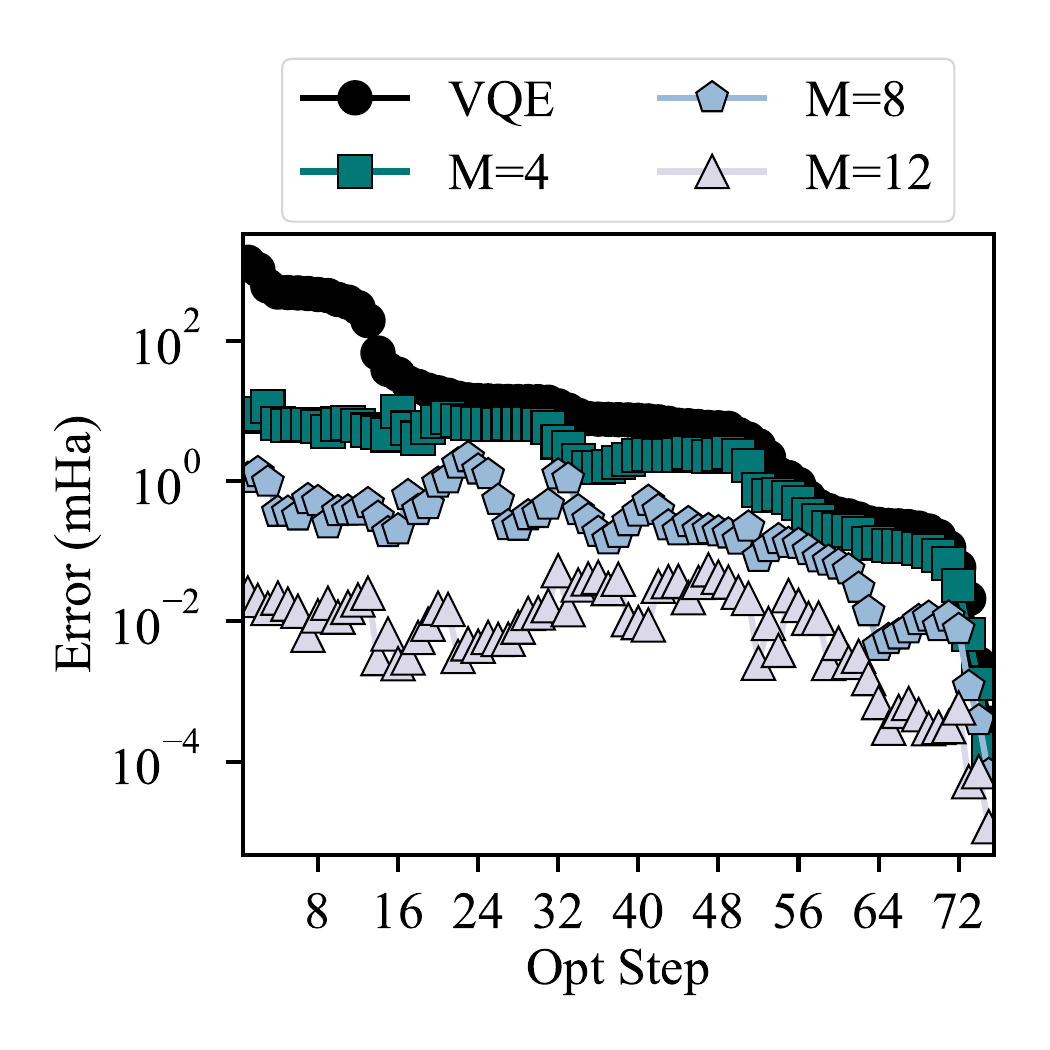}}
    \caption[]{(a) The error in the energy relative to FCI produced by 30 randomly-initialized UCC circuits for $\mathrm{H}_2$ using the 6-31G basis. The black dots indicate the error in the fully optimized VQE energies, arranged in ascending order. Four circuits obtained energies at or near the global minimum of the circuit, and the remaining circuits ended the optimization process at local minima or saddle points. Applying the CSVQE algorithm to these circuits using 4 intermediate states, marked with green squares, slightly reduces the energy. Applying CSVQE with 8 states, plotted with blue-green pentagons, accomplishes significant error reduction, sometimes producing errors as low as 10-100 $\mu$Ha even for circuits that obtained local minima. Applying CSVQE using 12 states achieves energies at parity with the global minimum for nearly all circuits. The scale is logarithmic above the break and linear below. We plot the convergence of the VQE and CSVQE energies as a function of the optimization step for the worst (trial index 30) and best (trial index 1) circuits in (b) and (c), respectively. In all CSVQE calculations we used 10,000 random combinations of mid-circuit states and report the lowest error obtained.}
    \label{fig:local_minima}
\end{figure*}

The reduction in error accomplished by the CSVQE algorithm ranges widely between both different molecules and different optimization steps of a given circuit. For all molecules studied here, CSVQE accomplishes significant error reduction by up to an order of magnitude early in the optimization process, often achieving parity with the error produced by the fully optimized circuit. \hl{However, we note that this is an absolute best case scenario and practical experiment likely cannot achieve the number of random samples required to obtain these results.} Furthermore, the error reduction produced by the CSVQE algorithm diminishes as the optimization proceeds, indicating that optimized circuits produce wavefunctions with energies that cannot be improved via superposition with mid-circuit states (at least within our numerical tests).

To gain insight into the distribution of the errors produced by these random combinations of states, rather than only considering the minima obtained, we plot in Figs.~\ref{fig:CSVQE_statistics} (a) and (b) the mean, minimum, and standard deviation of the CSVQE energies for the unoptimized and fully optimized $\mathrm{C}_2$ circuits as a function of $M$. We also plot in Figs.~\ref{fig:CSVQE_statistics} (c) and (d) histograms of the errors obtained for each circuit with $M=5$ using 10,000 random combinations of states. The data indicate that the errors are relatively flatly distributed around the average error, and the minimal errors occur as rare events below the main cluster of errors. Fortunately, the minimal errors are never significantly smaller than the bottom of the main distribution of errors, indicating that large amounts of sampling are not necessary to obtain good results. Furthermore, the width of the distribution narrows significantly as the circuit is optimized.

\section{Local Minima Analysis}\label{sec:local_minima}
One of the central challenges in implementing the VQE algorithm is handling local minima of the energy landscape. Classical optimizers often get trapped in these local minima, producing poor approximations of the groundstate energy~\cite{traps_2022,anuj_2022}. Here we study the ability of the CSVQE algorithm to produce accurate energies even when the VQE optimization terminates at a local minima.

We use the $\mathrm{H}_2$ molecule as a test case, employing the 6-31G basis set and the experimental geometry~\cite{CCCBDB}, using the UCC ansatz described above including all doubles operators with $N_{\text{WF}}=30,000$. We construct 30 circuits with random initial parameters and plot the error in the energy of the optimized VQE wavefunctions in black dots in Fig.~\ref{fig:local_minima} (a). Four of these circuits obtain energies at or near the true groundstate energy, but the remaining circuits obtained local minima (or possibly saddle points), producing energies on the order of tens or hundreds of milliHartree above the true groundstate energy. We also apply the CSVQE algorithm to each of these circuits using 10,000 random samples of $M=4,\,8,\mathrm{ and } \,12$ mid-circuit states and plot the resulting groundstate energy errors. The $M=4$ case somewhat reduces the error produced by the circuits trapped at local minima, but the $M=8$ and $M=12$ cases produce energies comparable to the global minima.  This result indicates that the CSVQE algorithm has significant potential as a powerful tool for circumventing local minima in VQE circuit optimization.

We also consider the performance of the CSVQE algorithm throughout the optimization of both the best (trial index 1) and worst (trial index 30) of these randomly initialized circuits. We plot the VQE and CSVQE energies of circuit 1 and circuit 30 as a function of the optimization step in Fig.~\ref{fig:local_minima} (b) and (c), respectively. For the energy before optimization, the initial error of circuit 30 is roughly 1.5 Ha, an astronomical error compared to any other reported in this work. Despite this enormous error, the CSVQE algorithm still produces energies with errors on the order of 10 mHa using only four mid-circuit states, or even on the scale of $\mu$Ha using twelve mid-circuit states. Although this is a relatively small circuit for a relatively simple system, it is remarkable that CSVQE can produce energies at or near the true ground state energy even for randomized circuits. The result of applying CSVQE to the best circuit is quite similar. For the unoptimized version of circuit 1, the error starts at the order of hundreds of mHa, but the CSVQE algorithm produces errors on the scale of 10 $\mu$Ha to 10 mHa, depending on the number of mid-circuit states used. The difference from circuit 30 is that the VQE and CSVQE energies converge as the circuit is optimized to obtain the true groundstate minimum.

\section{Conclusions}\label{sec:conclusion}
In this work we devised a modification of the VQE algorithm, dubbed circuit-subspace VQE, that utilizes quantum spaces generated by sampling mid-circuit states from VQE circuits. We applied the CSVQE algorithm to UCC circuits for a range of molecules, studying the reduction in error that can be accomplished with the technique at each step in the optimization process. We found that the CSVQE algorithm accomplished significant reduction in error \hl{in the best case} when applied to partially optimized circuits, but the advantage gained is minimal for fully optimized circuits.  However, due to shot noise, hardware noise and difficult optimization landscapes, optimization on quantum hardware is likely to be less than ideal and the tools laid out here provide another route to improving accuracy in such cases.

The CSVQE algorithm also shows significant promise for addressing the challenge of local minima in the optimization of VQE circuits. We showed that the CSVQE algorithm is able to produce energies comparable to the global minimum when applied to circuits for which the optimization process obtains a local minimum. Although this result was obtained for a relatively small circuit, we believe it indicates that the CSVQE algorithm can be applied to large circuits to diagnose cases in which the optimizer is stuck in a local minima. These promising results merit additional study.
     
The utility of this algorithm is likely highly ansatz-dependent, and further investigation of its utility beyond the UCC ansatz and quantum chemistry problems is justified. One important shortcoming of the UCC ansatz utilized in this work is that a large portion of the UCC parameters are quite small, leading to a circuit-generated subspace containing many nearly-identical wavefunctions. The CSVQE algorithm will likely perform significantly better on more efficient ansatze in which each gate meaningfully changes the wavefunction. This would also significantly lessen the challenge of designing efficient state selection strategies for generating useful subspaces. \hl{Another approach to easing the difficulty of choosing a useful subspace is to employ some measure of the overlap between the intermediate states as a selection criteria. The decoupling of the VQE optimization and the subspace calculations is a further area in which the algorithm could be improved, but we leave the question of how best to integrate circuit optimization with subspace calculations to future work.}

\hl{We note that our approach has standard scaling with similar Krylov method techniques that require matrix elements to be measured on quantum hardware. However, the utility of the algorithm heavily depends on the quality of the optimization of the circuit, as demonstrated in the different case studied in this work. This is certainly of interest, as optimization on quantum hardware is resource intensive and will likely remain resource intensive even as quantum hardware improves. The above considerations imply that this algorithm may benefit significantly from integration with existing techniques}~\cite{Anastasiou2022,Tang2021,PhysRevLett.131.200601}.

\textit{Acknowledgements}.--We are grateful for support from NASA Ames Research Center. 
We acknowledge funding from the NASA ARMD Transformational Tools and Technology (TTT) Project. 
W.M. and N.T. acknowledge funding from the NASA ARMD Transformational Tools and Technology (TTT) Project.
This material is based upon work supported by the U.S. Department of Energy, Office of Science, National Quantum Information Science Research Centers, Superconducting Quantum Materials and Systems Center (SQMS) under contract No. DE-AC02-07CH11359 (N.M.T.). 
M.R.H. participated in the Feynman Quantum Academy internship program.
Y.S., K.K., and R.V.B. acknowledge funding from the U.S. Department of Energy (DOE) under Contract No. DE-AC0205CH11231, through the Office of Science, Office of Advanced Scientific Computing Research (ASCR) Exploratory Research for Extreme-Scale Science.
This research used resources of the National Energy Research
Scientific Computing Center, a DOE Office of Science User Facility supported by the Office of Science of the U.S. Department of Energy under Contract No. DE-AC02-05CH11231 using NERSC award ASCR-ERCAP0024469.


%

\end{document}